\documentclass[aps,prb,amsmath,amssymb,twocolumn]{revtex4}
\usepackage{graphicx}
\usepackage{hyperref}

\begin{document}

\title{Evanescent states in 2D electron systems with spin-orbit interaction and spin-dependent transmission through a barrier}

\author{Vladimir~A.~Sablikov and Yurii~Ya.~Tkach}

\affiliation{Kotel'nikov Institute of Radio Engineering and Electronics,
Russian Academy of Sciences, Fryazino, Moscow District, 141190,
Russia}

\begin{abstract}
We find that the total spectrum of electron states in a bounded 2D electron gas with spin-orbit interaction contains two types of evanescent states lying in different energy ranges. The first-type states fill in a gap, which opens in the band of propagating spin-splitted states if tangential momentum is nonzero. They are described by a pure imaginary wavevector. The states of second type lie in the forbidden band. They are described by a complex wavevector. These states give rise to unusual features of the electron transmission through a lateral potential barrier with spin-orbit interaction, such as an oscillatory dependence of the tunneling coefficient on the barrier width and electron energy. But of most interest is the spin polarization of an unpolarized incident electron flow. Particularly, the transmitted electron current acquires spin polarization even if the distribution function of incident electrons is symmetric with respect to the transverse momentum. The polarization efficiency is an oscillatory function of the barrier width. Spin filtering is most effective, if the Fermi energy is close to the barrier height.
\end{abstract}

\maketitle

\section{Introduction}

The spin-orbit interaction (SOI) in low-dimensional structures attracts a great deal of interest since it opens up the possibility to manipulate the electron spin in nonmagnetic structures using electrical means.~\cite{Awschalom,Zutic} In this view, semiconductor heterostructures with 2D electrons are very promising since the Rashba SOI is effectively controlled~\cite{Nitta,Matsuyama,Schmult} by varying applied bias or gate voltages. In recent years, predominant interest was paid to effects appearing when the SOI modifies propagating electron modes with energy above the conduction band bottom. Suffice it to mention the spin-Hall effect,~\cite{Sinova,Chalaev,Schliemann} or spin manipulation in strained semiconductors~\cite{Kato}. In this paper we show that interesting effects of the SOI arise also when the electron energy is lower than or near to the conduction band bottom and evanescent states are involved. These states determine electron tunneling. They are important in 2D structures with laterally inhomogeneous potential landscape. We find that such structures can effectively polarize the transmitted electron current.

3D tunnel structures, in which spin polarization arises due to the SOI, were considered in a number of recent works. Zakharova et al~\cite{Zakharova} studied the interband tunneling, Voskoboynikov et al~\cite{Voskoboynikov} considered a tunnel structures with the Rashba SOI at the interfaces. In these cases the electron flow acquires a spin polarization if the structure is asymmetric. In symmetric tunnel structures the spin polarization arises, if the barrier material is noncentrosymmetrical.~\cite{Perel} The polarization mechanism, proposed by Perel' et al~\cite{Perel,Tarasenko}, consists in a spin-dependent renormalization of the electron effective mass owing to the Dresselhaus SOI in the barrier. However, all these structures have a common property restricting their capability to generate spin polarization. The polarization is absent if the electron current is perpendicular to the barrier. In other words, for the spin polarization to appear the momentum distribution function of incident electrons must be asymmetric with respect to the momentum component parallel to the barrier.

The effective mass renormalization occurs if the Hamiltonian of the SOI is quadratic in longitudinal momentum. However, the dispersion relation of electrons in the presence of the SOI is generally much more complicated and therefore a more careful analysis of the complex band structure and evanescent states should be carried out to study the spin-dependent tunneling. In the 3D case, such calculations were recently carried out for some specific materials and qualitatively new features were found.~\cite{Mishra,Sandu,Wang}

2D tunnel structures are scantily studied to date. In particular, as far as we know, even the complex band structure of 2D electrons was not explored. Though the presence of evanescent states is obvious, only a few of works touched upon these modes. Usaj, Reynoso and Balseiro~\cite{Usaj,Reynoso} attracted evanescent states to study the electron scattering at the edges of 2D samples, but the total spectrum of evanescent states was not considered. The importance of evanescent modes in quasi-one-dimensional systems in the presence of the SOI was pointed out in a number of works.~\cite{Governale,Streda,Lee,Serra}

Khodas, Shekter, and Finkel'stein~\cite{Khodas,Shekhter} studied the electron beam propagation in 2D electron gas with spatially inhomogeneous SOI. They considered the transmission through a strip, in which the SOI strength differs from that in the rest of the 2D electron gas, to show that an initially unpolarized beam splits into two beams with different spin polarizations propagating in different directions. The consideration was restricted by propagating states since only the case of uniform potential landscape was studied. Spin-dependent reflection of electrons from a lateral barrier in 2D system was observed in InSb/InAlSb heterostuctures~\cite{Chen} and described theoretically in Ref.~\onlinecite{Govorov}. Silvestrov and Mishchenko~\cite{Silvestrov} demonstrated the possibility to achieve spin-polarized currents in a 2D system with smooth potential barrier and spatially-uniform SOI by considering propagating modes within semiclassical approach.

In this paper we show that strong polarization effect appears in 2D structures when electrons pass through a lateral potential barrier, in which the SOI is stronger than in the outside 2D electron gas regions (reservoirs). The polarization arises even if the electric current is normal to the barrier. The highest polarization is attained when the electron energy is close to the conduction band bottom in the barrier. In this case the fact becomes important that the SOI effectively splits the barrier height so that some part of electrons passes through the barrier via propagating states while others do this via the evanescent modes. Since the spin and orbital degrees of freedom are coupled, rather strong spin filtering occurs.

We study the total spectrum of electron states in 2D bounded system with SOI to find that there are two type of evanescent states. The first-type states are characterized by an imaginary longitudinal wavevector. They fill in a gap in the propagating state spectrum. The states of the second type lie in the forbidden gap. They are described by a complex wavevector. The electron tunneling through a lateral barrier with SOI via these evanescent states exhibits unusual features, such as an oscillatory behavior of the transmission coefficient with the barrier width. But of most interest is the spin polarization of the electron current. The polarization efficiency is high enough even if the distribution function of incident electrons is symmetric with respect to the transverse momentum. We explore the polarization efficiency in a wide range of electron energy to find that most effective spin filtering occurs if the Fermi energy is close to the barrier height.

The paper is organized as follows. In Sec.~\ref{ComplexStructure} we describe the complex band structure and total spectrum of electron states. Sec.~\ref{Tunneling} is devoted to the tunneling through a barrier with SOI. In Sec.~\ref{Polarization} the electric and spin currents through the barrier with SOI  are considered for a wide range of the Fermi energy in the reservoirs. The results obtained for the Rashba SOI are generalized to the Dresselhaus SOI in Sec.~\ref{Dresselhaus}. We summarize and conclude in Sec.~\ref{Conclude}.

\section{Complex band structure of a 2D electron system with SOI}
\label{ComplexStructure}

We start by considering the total spectrum of propagating and evanescent electron states in 2D electron gas with the Rashba SOI. The Hamiltonian is~\cite{Rashba}
\begin{equation}
H_R=\dfrac{\hbar^{2}}{2m}(p_{x}^{2}+p_{y}^{2})+
\frac{\alpha}{\hbar}(p_{y}\sigma_{x}-p_{x}\sigma_{y})\,,
\label{HamiltR}
\end{equation}
where $\alpha$ is the SOI constant, $\sigma_x,\sigma_y$ are Pauli matrices. The eigenfunctions are
\begin{equation}
\psi_{\mathbf{k},s}=C e^{i(k_x x+k_y y)}
\binom{\chi_s(\mathbf{k})}{1}\,,
\label{psi}
\end{equation}
where $s=+,-$ stands for spin states, $\mathbf{k}=(k_x,k_y)$, $C$ is a constant. The eigenenergy $\varepsilon_{\mathbf{k},s}$ and the spin function $\chi_s(\mathbf{k})$ are defined by following equations:
\begin{equation}
\left\{
\begin{array}{rl}
(\zeta_{\mathbf{k},s}-k^2)\chi_s(\mathbf{k}) - 2a(k_y+ik_x)&=0 \\
- 2a(k_y-ik_x)\chi_s(\mathbf{k}) + (\zeta_{\mathbf{k},s}-k^2) &=0\;,
\end{array}
\right.
\label{zeta-chi}
\end{equation}
where $a=m\alpha/\hbar^2$ is a characteristic wavevector of SOI, $k^2=k_x^2+k_y^2$ and $\zeta_{\mathbf{k},s}$ is the normalized eigenenergy:
\begin{equation*}
\zeta_{\mathbf{k},s}=\dfrac{2m\varepsilon_{\mathbf{k},s}}{\hbar^2}\,.
\end{equation*}
Using Eqs~(\ref{zeta-chi}) one obtains the dispersion equation
\begin{equation}
 (\zeta_{\mathbf{k},s}-k^2)^2-4a^2k^2=0
\label{zeta}
\end{equation}
and the spin function
\begin{equation}
 \chi_s(\mathbf{k})=2a\dfrac{k_y+ik_x}{\zeta_{\mathbf{k},s}-k^2}\,.
\label{chi}
\end{equation}

Let us analyze the dispersion equation. To be specific, assume that the system under consideration is infinite in $y$ direction and has a boundary in $x$ direction, i.e. the system is semi-infinite or finite in the $x$ direction. In this case, the $y$ component of the wavevector $k_y$ is real, while $k_x$ is generally complex, $k_x=k'_x+ik''_x$. Dividing the real and imaginary parts of Eq.~(\ref{zeta}) we obtain an equation set determining the energy $\zeta$ as a function of $k'_x,k''_x,k_y$ and trajectories in the ($k'_x,k''_x$) plane, along which $\zeta(k'_x,k''_x,k_y)$ is real:
\begin{align}
\label{dis1}
&(\zeta-k'^2+k''^2)^2=4a^2(k'^2-k''^2)+4k'^2k''^2\,,\\ 
\label{dis2}
&(\zeta-k'^2+k''^2)k'k''=-2a^2k'k''\,,
\end{align}
where $k'+ik''=\sqrt{(k'_x+ik''_x)^2+k_y^2}$. 

\begin{figure}[h,t]
\includegraphics[width=0.9\linewidth]{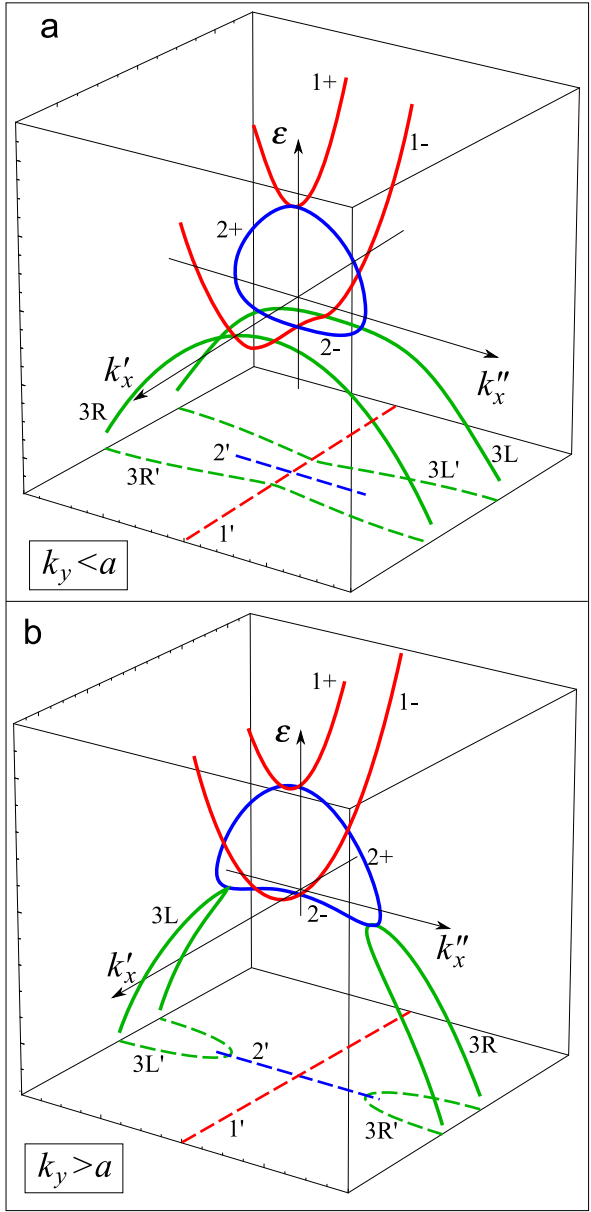}
\caption{(color online). Total spectrum of 2D electron gas with SOI. Solid lines $1+$ and $1-$ are spin-splitted propagating modes of branch 1; lines $2+$ and $2-$ are evanescent states, branch 2; lines $3L$, $3R$ are evanescent states in the forbidden gap, branch 3. Dashed lines $1'$, $2'$ and $3L'$, $3R'$ are the trajectories corresponding to these branches on the complex plane ($k'_x,k''_x$).}
\label{f_spectrum}
\end{figure}

Eq.~(\ref{dis2}) possesses a solution in the following cases: i)~$k'=0$,~ ii)~$k''=0$ and iii)~$\zeta-k'^2+k''^2=-2a^2$ for $k',k''\neq 0$. The first case contradicts to Eq.~(\ref{dis1}) and hence cannot be realized. The second case generates two branches of the solution, which can be found after dividing the real and imaginary parts of Eq.~(\ref{dis1}):

1. $k''_x=0$,
\begin{equation}
\begin{split}
 \zeta_{\mathbf{k},s}&=-a^2+\left(a\pm \sqrt{k_y^2+{k'_x}^2}\right)^2\,,\\
 \chi_s(\mathbf{k})&=\pm\dfrac{k_y+ik'_x}{\sqrt{k_y^2+{k'_x}^2}}\,;
\end{split}
\label{branch1}
\end{equation}

2. $k'_x=0$, $|k''_x|<|k_y|$,
\begin{equation}
\begin{split}
 \zeta_{\mathbf{k},s}&=-a^2+\left(a\pm \sqrt{k_y^2-{k''_x}^2}\right)^2\,,\\
 \chi_s(\mathbf{k})&=\pm\dfrac{k_y-k''_x}{\sqrt{k_y^2-{k''_x}^2}}\,.
\end{split}
\label{branch2}
\end{equation}

\noindent The third case gives one further branch: 

3. This branch is defined for $k'_x,k''_x$ belonging to the folowing trajectory
\begin{equation}
 {k'_x}^2{k''_x}^2+a^2(k_y^2+{k'_x}^2-{k''_x}^2)-a^4=0
\end{equation}
in the complex plane ($k'_x,k''_x$). The eigenenergy and the spin function are
\begin{equation}
\label{branch3}
\begin{split}
 \zeta_{\mathbf{k},s}&=-a^2-\dfrac{{k'_x}^2{k''_x}^2}{a^2}\,,\\
 \chi_s(\mathbf{k})&=-a\dfrac{k_y-k''_x+ik'_x}{a^2+ik'_xk''_x}\,.
\end{split}
\end{equation}


The total spectrum is shown schematically in Fig.~\ref{f_spectrum} where the energy $\varepsilon_{\mathbf{k},s}$ is presented as a function of $k'_x,k''_x$ for given transverse momentum $k_y$. The form of all three branches is different in the cases $|k_y|<a$ and $|k_y|>a$. 

Branch 1 describes the propagating states with real $k_x$. This branch is splitted by spin. The energy gap, which opens at $k_x = 0$ for $k_y \neq 0$, depends on $|k_y|$.

Branch 2 describes purely decaying evanescent states defined on the imaginary $k_x$ axis. This branch connects the propagating state branches along the imaginary axis. The spectrum of branch 2 is also splitted by spin. The energy minimum $\varepsilon_m = -\hbar a^2/(2m)$ is common for branches 1 and 2. The minimum is attained on branch 1, if $|k_y|<a$, or on branch 2 in the opposed case.

\begin{figure}[h,t]
\includegraphics[width=0.9\linewidth]{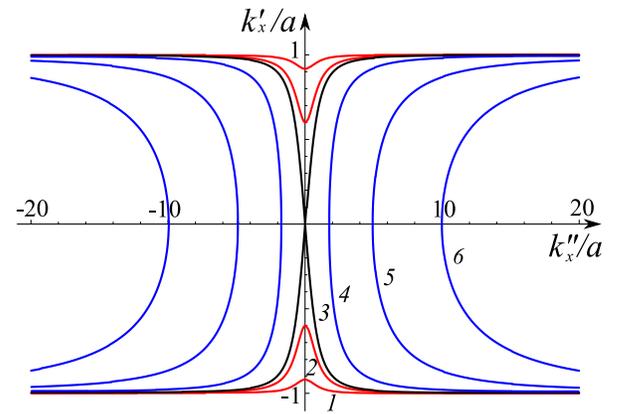}
\caption{(color online). Real-energy trajectories along which the ``oscillating'' evanescent states are defined. $k_y$ is fixed for each line: lines \textit{1-6} correspond to $k_y/a = 0.4, 0.8, 1.0, 2.0, 5.0, 10.0$.}
\label{f_trajectories}
\end{figure}

Branch 3 corresponds to evanescent states in the forbidden gap, $\varepsilon_{\mathbf{k},s}<\varepsilon_m$. They are described by a complex longitudinal wavevector and therefore can be named ``oscillating'' evanescent states. The trajectories, along which these states are defined, obey the following equation:
\begin{equation}
(a^2 - {k'_x}^2)(a^2 + {k''_x}^2) = a^2 k_y^2 \,.
\label{trajectories}
\end{equation}
They are shown in Fig.~\ref{f_trajectories}. There are trajectories of two types. If $|k_y|<a$, the trajectories (lines 1, 2) intersect the imaginary axis $k''_x$. In the vicinity of the intersection point, $|k''_x| \ll |k'_x|$ and hence the wavefunction oscillates with the distance faster than decreases. If $|k_y|>a$ the trajectories (lines 4-6) intersect the $k'_x$ axis.

It is seen that there are four states for any energy, in accordance with the number of the degrees of freedom of the system. At a given energy all the states are distinguished by the wavevector components $k'_x,k_x''$ and the spin function. The propagating states and the evanescent states of branch 2 have two subbranches divided by the energy, while the evanescent states of branch 3 are splitted in the complex momentum plane.

Though the wavefunctions of the third branch are complex, they do not carry the current. Using the Hamiltonian (\ref{HamiltR}) one obtains the following expression for the particle current in the state 
\begin{equation}
\label{psi_spinor}
\psi_s=\binom{\psi_1}{\psi_2}\,,
\end{equation}
\begin{equation}
\label{current}
\begin{split}
\mathbf{j}_s&=  \dfrac{i\hbar}{2m}\left(\psi_1\nabla\psi^*_1-\psi^*_1\nabla\psi_1+\psi_2\nabla\psi^*_2-\psi^*_2\nabla\psi_2\right)\\ & -\dfrac{i\alpha}{\hbar}\left(\psi_1\psi^*_2-\psi^*_1\psi_2\right)\mathbf{e}_x +\dfrac{\alpha}{\hbar}\left(\psi^*_1\psi_2+\psi_1\psi^*_2\right)\mathbf{e}_y .
\end{split}
\end{equation}
The wavefunctions of branch 3, calculated with using Eqs~(\ref{psi}) and (\ref{branch3}), are easily seen to turn the $x$ component of the current (\ref{current}) to zero while $y$ component is nonzero.

\section{Tunneling currents through a barrier with SOI}
\label{Tunneling}

In this section we study the electron tunneling via the ``oscillating'' evanescent states of branch 3. Consider a 2D electron gas without SOI divided into two semiplanes (reservoirs) by a rectangular barrier, in which the SOI is present. The barrier height is $U$ and the width is $d$.

Let us calculate the transmission probability for electrons incident on the barrier from the left reservoir. We use the presentation of the wavefunctions in the reservoirs in the basis of eigenstates $|s\rangle$ of the $\sigma_z$ matrix. The basis functions are $|k_x,k_y,s\rangle$, where $k_x$ and $k_y$ are the wavevector components in the reservoirs, $s=\uparrow,\downarrow$.

In the left reservoir ($x<0$) the wavefunction is
\begin{equation}
 |\psi^{(L)}_{k_x,k_y,s}\rangle =| k_x,k_y,s\rangle + \sum_{s'}r_{s,s'}|-k_x,k_y,s'\rangle\;,
\label{Lwave}
\end{equation}
where $|k_x,k_y,s\rangle$ is the state vector of incident electrons. The wavefunction of transmitted electrons ($x>d$) is 
\begin{equation}
 |\psi^{(R)}_{k_x,k_y,s}\rangle = \sum_{s'}t_{s,s'}|k_x,k_y,s'\rangle\;.
\label{Rwave}
\end{equation}
Here $r_{s,s'}$ and $t_{s,s'}$ are reflection and transmission matrices.

The wavefunction in the barrier is expanded in the eigenstates~(\ref{psi}) of the Hamiltonian~(\ref{HamiltR})
\begin{equation}
|\psi^{(B)}_{k_x,k_y,s}\rangle =\sum_{r,r'=+,-} b_{r,r'}^s|rK'_x,r'K''_x,k_y\rangle \,,
\label{Bwave}
\end{equation}
where $K_x=K'_x+iK''_x$ denotes the complex wavevector in the barrier; $r,r'=\pm$ are indexes labeling all four evanescent states in the barrier. They are described by the spinors~(\ref{psi}), in which the real and imaginary parts of $K_x$ should be taken with different signs.

The matrices $r_{s,s'}$, $t_{s,s'}$ and $b_{r,r'}^s$ are determined by an equation set, which follows from the boundary conditions at the interfaces of the barrier and 2D electron reservoirs.

Boundary conditions for wavefunctions are obtained in a standard way by integrating the Schr\"odinger equation over the infinitesimal vicinity of the boundary. These conditions are well known for a boundary between regions with different strength of SOI~\cite{Molenkamp,Khodas}. In our case it is necessary to take into account that the lateral potential step at the boundary also contributes to the SOI. In the transition region, where the potential $U(x)$ varies with $x$, the following additional term should be added to the Hamiltonian~(\ref{HamiltR})
\begin{equation}
\label{boundarySOI}
H^b_{so}=\dfrac{\gamma}{\hbar}\dfrac{dU}{dx}\sigma_z p_y\,,
\end{equation}
where $\gamma$ is SOI constant connected with $\alpha$ by $\alpha=-e\gamma F_z$, $F_z$ being the electric field perpendicular to the 2D layer. This term having been integrated over the transition region gives a finite contribution to the boundary conditions. Finally one obtains the following equations for the spinor~(\ref{psi_spinor}) components:
\begin{align}
\label{boundary_cond1}
&\psi_1|_{-0}^{+0} = \psi_2|_{-0}^{+0} =0 \,;\\
\label{boundary_cond2}
&\dfrac{1}{m(x)}\left[\dfrac{\partial \psi_1}{\partial x} \pm \beta(x) k_y\psi_1- a(x) \psi_2\right]_{-0}^{+0}=0\\ 
\label{boundary_cond3}
&\dfrac{1}{m(x)}\left[\dfrac{\partial \psi_2}{\partial x} \mp \beta(x)k_y \psi_2 + a(x) \psi_1\right]_{-0}^{+0}=0\,,
\end{align}
where the parameter 
\begin{equation}
\beta= a \dfrac{2U}{eF_z}\,
\label{beta}
\end{equation}
describes the Rashba SOI caused by the in-plane field. The upper and lower signs in Eqs~(\ref{boundary_cond2}),(\ref{boundary_cond3}) correspond to the boundaries at which $U(x)$ increases or decreases with $x$.

Applying these boundary conditions to the system under consideration we put $a=0$ in the 2D electron reservoirs and keep $a\ne 0$ in the barrier. For simplicity, we ignore the difference in the effective masses of electrons in the barrier and reservoirs.

In addition, the wavevectors in the barrier and the reservoirs should be matched. The tangential components $k_y$ must be equal. The relation of the normal components is determined by equaling the energy $E$ of an incident electron to the electron energy in the barrier 
\begin{equation}
 E(k_x,k_y) = U + \varepsilon(K_x,k_y,s)\;,
\end{equation}
where $\varepsilon(K_x,k_y,s)$ is defined by Eqs~(\ref{branch1}),(\ref{branch2}),(\ref{branch3}) in accordance with the energy spectrum branch (1, 2 or 3), which is considered. In the present section we restrict ourselves by branch 3.

Finally one obtains two equation sets for the cases of the incident spin directed up and down. Each equation set contains eight equations. Dropping elementary calculations and combersome expressions for matrices $t_{s,s'}$, $r_{s,s'}$ and $b_{r,r'}^s$, we turn directly to main results.

First, note that the $t_{s,s'}$ matrix obeys the following symmetry relations:
\begin{equation}
\label{symmetry}
\begin{split}
t_{\uparrow\uparrow}(K_x,k_y)=&t_{\downarrow\downarrow}(K_x,-k_y)\,,\\
t_{\uparrow\downarrow}(K_x,k_y)=&-t_{\downarrow\uparrow}(K_x,-k_y)\,.
\end{split}
\end{equation}

The particle current through the barrier is
\begin{equation}
\label{current_K}
 j_{k_x,k_y,s}=\dfrac{\hbar k_x}{m}T_s(k_x,k_y)\;,
\end{equation}
where $T_s$ is the transmission coefficient, $T_s=\sum_{s'}|t_{s,s'}|^2$, which has a meaning of the probability for an electron to tunnel with any spin in the final state. The peculiarity of the tunneling in the presence of the SOI consists in the involvement of four interfering states with different spin structure.

\begin{figure}[h,t]
\includegraphics[width=0.9\linewidth]{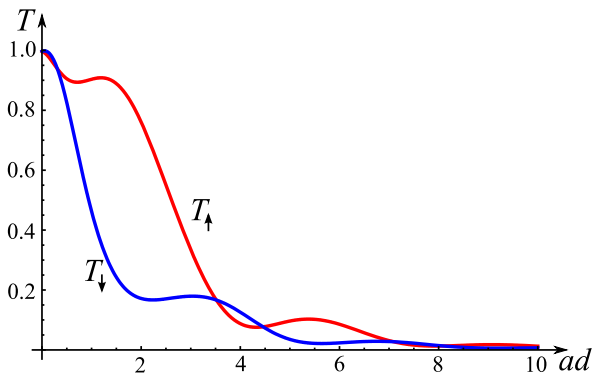}
\caption{(color online). Tunneling coefficients for incident electrons with spin up and down as functions of the barrier width. $U=2E_{so}$, $E=0.95E_{so}$, $k_y=0.49 a$ and $\beta=0.1$.}
\label{tunnel_oscillat}
\end{figure}

One of unusual consequences of this fact is an oscillatory dependence of the tunneling coefficient on the barrier width. This feature is demonstrated in Fig.~\ref{tunnel_oscillat}. The oscillations exist when the electron energy is close to the top of the third branch: $U-E_{so}-E\ll E_{so}$, where $E_{so}=\hbar^2a^2/(2m)$ is a characteristic energy scale of SOI. The oscillations fade away as the energy decreases deep into the forbidden band, because $K''_x$ exceeds $K'_x$. The oscillations disappear also if the tangential momentum $k_y$ goes to zero.

Fig.~\ref{tunnel_oscillat} shows clearly that the barrier filters incident electrons by spin. This process depends evidently on the incident angle and the energy of electrons. Consider the spin polarization of transmitted electrons in more detail. Let an unpolarized electron flow with the wavevector $(k_x,k_y)$ is incident on the barrier. The flow consists of two components $|k_x,k_y,\uparrow\rangle$ and $|k_x,k_y,\downarrow\rangle$ with opposed spins. The transmitted flow acquires spin polarization. The spin density of transmitted electrons is
\begin{equation}
\vec{\mu}_{k_x,k_y}=\dfrac{\hbar}{2}\sum_{s=\pm 1}\langle \psi_{k_x,k_y,s}^{(R)}|\hat{\vec{\sigma}}|\psi_{k_x,k_y,s}^{(R)}\rangle\;.
\label{mu_k-s}
\end{equation}
With using Eq.~(\ref{Rwave}), the spin density takes the form:
\begin{equation}
\vec{\mu}_{k_x,k_y}\! =\! \dfrac{\hbar}{2}|C|^2 \! \sum_{s,s'}\!\left[(\mathbf{e}_x\!-\! i \mathbf{e}_y s') t^*_{s,s'}t_{s,-s'}+ \mathbf{e}_z s'|t_{s,s'}|^2 \right],
\label{mu_k-s1}
\end{equation}
where $|C|^2$ is a normalization constant.

\begin{figure}[h,t]
\includegraphics[width=0.9\linewidth]{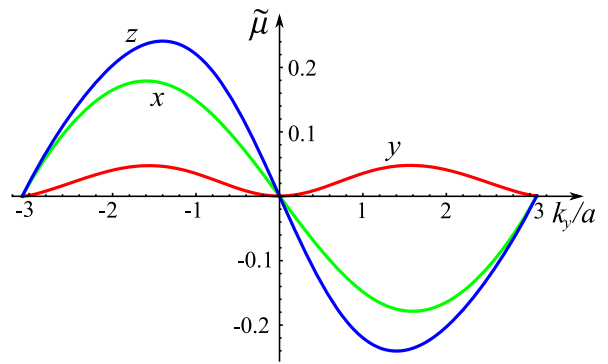}
\caption{(color online). Spin polarization of transmitted electrons, $\tilde{\mu}=2\mu_{x,y,z}/\hbar |C|^2$. The curves are marked by letters corresponding to the polarization components. The incident electron energy $E=9.5~E_{so}$, the barrier height $U=11~E_{so}$, the barrier width $d=a$, the lateral SOI parameter $\beta=0.1$.}
\label{tunnel_polariz}
\end{figure}

The components of the spin polarization are shown in Fig.~\ref{tunnel_polariz} as functions of the transverse momentum $k_y$ for a given energy of incident electrons. The main property of the acquired polarization is that the $x$ and $z$ components of the spin polarization are odd functions of $k_y$, while the $y$ component is an even function of $k_y$. This property is independent of the electron energy, the barrier height, the SOI strength, and thus is universal for the Rashba SOI. Two consequences follow from this fact. First, the unpolarized electron follow acquires spin polarization even if the distribution function of the incident electrons is symmetric with respect to the tangential momentum direction, the polarization being directed parallel to the barrier. Second, if the distribution function is not symmetric about $k_y$, the spin polarization arises also in the $x$ and $z$ directions.

Detail analysis shows that the spin polarization is caused mainly by the SOI in the barrier, as it is described by the Hamiltonian~(\ref{HamiltR}). The SOI at the boundaries of the barrier (described by Eq.~(\ref{boundarySOI})) does not essentially affect the results if the parameter $\beta$ defined by Eq.~(\ref{beta}) is small. This case corresponds to realistic situation in experiments. Numerical estimations for InAs quantum well ($\alpha \sim 6\times 10^{-9}$~eVcm, $F_z\sim 10^5$~V/cm, $U\sim 20$~meV) give $\beta \sim 0.1$. If $\beta \gtrsim 1$, the boundary SOI changes the value of polarization, but main features (such as the symmetry relations, the oscillatory behavior of the tunneling coefficients) remain qualitatively similar.

\section{Spin polarization of electrons by a barrier with SOI}
\label{Polarization}

In this section we turn from the separate electron states to the total spin polarization produced by all electron states contributing to the current through a barrier with SOI. In addition, we consider a wide range of the incident electron energy to include all three branches of the electron spectrum. This allows one to find out conditions under which the electron current is polarized most effectively.

\begin{figure}[h,t]
\includegraphics[width=0.9\linewidth]{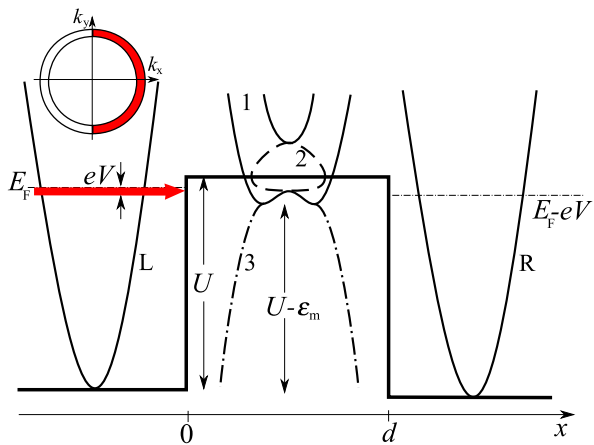}
\caption{(color online). A schematic view of the barrier with SOI. Lines L and R show the electron dispersion relations in the left and right reservoirs. Lines 1,2,3 image schematically corresponding branches of the dispersion relation in the barrier. The insertion is the $k_x,k_y$ space, in which the shaded region shows the electron states contributing to the current.}
\label{scheme}
\end{figure}

Consider an electron current directed normally to the barrier. For definiteness, let the current be caused by a voltage $V$ applied across the 2D electron reservoirs, as it is shown in Fig.~\ref{scheme}. The incident electron states, which contribute to the current, occupy an energy layer near the Fermi level. They are located in a semi-ring in the $k_x,k_y$ plane, shown in the insertion. The particle and spin currents through the barrier are determined by summing over all these states. We carry out this calculation for various positions of the Fermi level $E_F$ relative to the barrier height $U$ to find the spin-polarization efficiency as a function of $E_F$.

For simplicity suppose that the voltage is small compared to other energies, $eV\ll U, E_F, E_{so}$. This simplification allows one to restrict the summation by the integration in $\mathbf{k}$ space over the azimuthal angle at a given energy $E$. The integration is conveniently to carry out over $k_y$, but in doing this the fact should be taken into account that the set of branches, which must be used at given $k_y$ and $E$, can change with varying $k_y$. This happens because the gap $\Delta \varepsilon_{12}$ between spin-splitted subbranches of propagating modes (curves 1+ and 1- in Fig.~\ref{f_spectrum}) increases with $k_y$ ($\Delta \varepsilon_{12}=2\hbar^2a|k_y|/m$) and furthermore the form of the dispersion curves describing the propagating (1+, 1-) and evanescent (2+, 2-) modes changes. The switches between the actual branches occur when any extremum of the dispersion curves (which is a function of $k_y$) coinsides with the energy $E$. Physically this means that electrons incident on the barrier at different angles and in different spin states feel different effective barrier height.

\begin{figure}[h,t]
\includegraphics[width=1.0\linewidth]{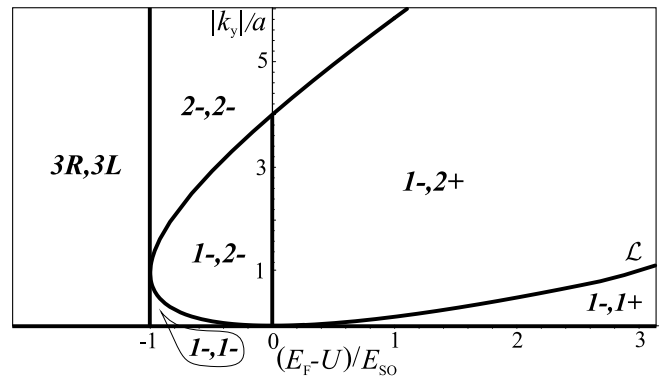}
\caption{The diagram of the spectrum branches accessible for electrons with energy $E$ and tangential momentum $k_y$. Thick lines divide the $(E,k_y)$ plane into 6 regions, in which corresponding branches are denoted by numbers 1,2,3 and chirality indexes $+,-$ in accordance with Fig.~\ref{f_spectrum}. Line $\mathcal L$ is defined by equation $ k_y/a = (\sqrt{(E-U-E_{so})/E_{so} +1} \pm 1)^2$}
\label{diagram}
\end{figure}

Fig.~\ref{diagram} presents a diagram showing which branches of the dispersion relation are accessible for electrons with given $E$ and $k_y$. Within each region bounded by thick lines, there are two different branches with two fundamental eigenfunctions on each ones or one branch with four solutions.

For each region of the diagram the wavefunctions are determined in the same way as in Sec.~\ref{Tunneling}. The only difference is that the eigenfunctions of the branches specified in the diagram are to be used in Eq.~(\ref{Bwave}) instead of the wavefunctions of branch 3. As a result of these calculations the transmission $t_{s,s'}(E,k_y)$ and reflection $r_{s,s'}(E,k_y)$ matrices are obtained for the whole $(E,k_y)$ plane. Using $t_{s,s'}(E,k_y)$ we find the particle and spin currents in the right electron reservoir.

The particle current is found by the summation of Eq.~(\ref{current_K}) over all states of incident electrons:
\begin{equation}
J=\dfrac{eV}{2\pi h}\int_{-k_F}^{k_F}\!\! dk_y \left(|t_{\uparrow \uparrow}|^2\!+\!|t_{\uparrow \downarrow}|^2\!+\!|t_{\downarrow \uparrow}|^2\!+\!|t_{\downarrow \downarrow}|^2\right)\,,
\end{equation}
where the integration symbol implies also the summation over all regions of the $(E,k_y)$ plane which fall within the $(-k_F,k_F)$ interval, with $k_F$ being the Fermi wavevector in the 2D reservoirs.

The transmitted spin current in the left reservoir is defined in a standard way~\cite{Rashba1,Shi} using the following expression for the current in a state $|k_x,k_y\rangle$:
\begin{equation}
J_{s,i}^j(k_x,k_y)= \dfrac{1}{2}\langle v_i \sigma_j + \sigma_j v_i \rangle\,,
\end{equation}
where $i=\{x, y\}$ denotes the velocity components in the plane, $j=\{x, y, z\}$ denotes the spin components in 3D space. In the case under consideration, the $x$ component of the total spin current is 
\begin{equation}
\label{spin_current}
J^j_s=\dfrac{eV}{2\pi h}\int_{-k_F}^{k_F}\!\! dk_y \!\left( 
\begin{array}{l}
\!2 \mathrm{Re} (t_{\uparrow \uparrow}t^*_{\uparrow \downarrow}\!+\!t_{\downarrow \downarrow}t^*_{\downarrow \uparrow})\\
\!2 \mathrm{Im} (-t_{\uparrow \uparrow}t^*_{\uparrow \downarrow}\!+\!t_{\downarrow \downarrow}t^*_{\downarrow \uparrow})\\
\!|t_{\uparrow \uparrow}|^2\!-\!|t_{\uparrow \downarrow}|^2\!+\!|t_{\downarrow \uparrow}|^2\!-\!|t_{\downarrow \downarrow}|^2
\end{array}
\!\!\right)\!,
\end{equation}
where three lines in the RHS correspond to the $x,y,z$ components of the spin polarization for the spin current directed along $x$ axis.

The efficiency of spin polarization is characterized by the ratio of the spin current to the particle current
\begin{equation}
\mathcal{P}_j = \dfrac{J^j_s}{J}\;.
\end{equation}

Using the symmetry relations~(\ref{symmetry}) and Eq.~(\ref{spin_current}) one finds that $x$ and $z$ components of the spin current are absent in the case of the Rashba SOI, and only the $y$ component is nonzero.  Of course, this is a consequence of the symmetry of the distribution function with respect to the sign of $k_y$. If the current had not be perpendicular to the barrier, the polarization would appear also in the $x$ and $z$ directions. The polarization efficiency turns out to be sufficiently high. The dependencies of the polarization on the Fermi energy and the barrier width are nontrivial because they reflect a complex structure of the electron spectrum. Below two most significant results are considered.

\begin{figure}[h,t]
\includegraphics[width=0.9\linewidth]{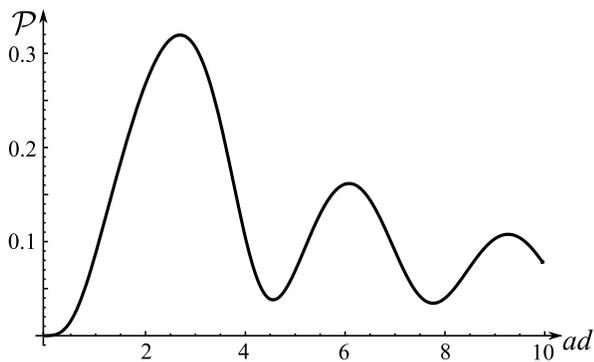}
\caption{Dependence of the polarization efficiency on the barrier width for the Fermi energy below $E_b$ for tunneling through branch 3. $U=5 E_{so}$, $E_F=3.5 E_{so}, \beta=0.1$.}
\label{polar-d_3}
\end{figure}

The dependence of $\mathcal{P}_y$ on the barrier width is shown in Fig.~\ref{polar-d_3} for energies below the top of the energy band where the evanescent states of branch 3 exist in the barrier, $E_F<U-E_{so}\equiv E_b$. The polarization efficiency is seen to oscillate with $d$ because of the interference of four oscillating evanescent modes. The effect is rather strong and becomes the stronger the closer is the energy to the band top. The oscillation period is of the order of $\pi/a$.

\begin{figure}[h,t]
\includegraphics[width=0.95\linewidth]{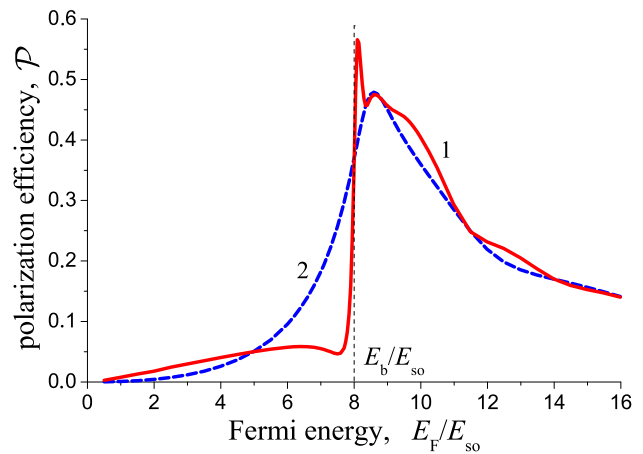}
\caption{(color online). Dependence of the polarization efficiency on the Fermi energy in 2D electron reservoirs. $U=9 E_{so}$ for $d=5/a$ (solid line 1) and $d=3/a$ (dashed line 2); $\beta=0.1$. Vertical dashed line marks the threshold energy $E_b=U-E_{so}$, which divides the evanescent states of branch 3 (on the left of $E_b$) from the states of branches 2 and 1.}
\label{polar-E}
\end{figure}

The dependence of $\mathcal{P}_y$ on the Fermi energy is presented in Fig.~\ref{polar-E} for a wide energy range including the energy both below and above the barrier. It is seen that there is a critical energy $E=8 E_{so}$, which coincides with the top energy of the third branch band, $E_b=U-E_{so}$, in the barrier. At energies well below $E_b$, the polarization efficiency $\mathcal{P}_y$ increases with the energy. In the vicinity of the threshold $E_b$ an oscillatory behavior appears as a result of the interference of four slowly decaying waves. Above this energy the transmission process goes via the propagating states of branch 1 and the evanescent states of branch 2. At $E_F$ slightly higher than $E_b$ the polarization efficiency attains the highest value. With further increasing the energy the efficiency $\mathcal{P}_y$ decreases. It is worth noting that the highest efficiency of the spin polarization slightly depends on the SOI strength, but the barrier width, at which this high polarization is attained, varies inversely with the SOI constant.

Fig.~\ref{polar-E} demonstrates also how $\mathcal{P}_y$ changes with the barrier width. Increasing $d$ above $\sim \pi/a$ leads to more pronounced oscillatory dependence of $\mathcal{P}_y$ on the energy caused by the four-wave interference.

\begin{figure}[h,t]
\includegraphics[width=0.9\linewidth]{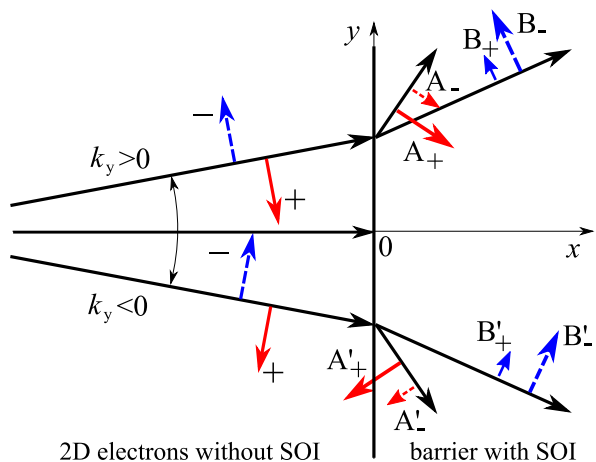}
\caption{(color online). Refraction and spin polarization of electron beams incident on the semiinfinite barrier with SOI. Solid arrows show the spin polarization in the case of positive chirality of the incident beam. Dashed arrows are the polarization for negative incident chirality. $A_{\pm}$ and $B_{\pm}$ are amplitudes for incident beam with $k_y>0$. The primed letters denote the same amplitudes for $k_y>0$.}
\label{birefringence}
\end{figure}

The physical mechanism, owing to which the polarization of normally incident electron current appears, is connected with the splitting of electron waves in the barrier because of the SOI. Let us consider first a simplified case of semiinfinite barrier region with SOI. Let the energy is high enough so that electrons occupy propagating states (branches 1+,1-). The incident electron states can be represented in terms of two chiral modes:
$$
\chi^{(0)}_{\pm}=\binom{\pm \chi}{1}\;, \qquad \chi=\dfrac{k_y+ik_x}{\sqrt{k_y^2+k_x^2}}\,.
$$
In the barrier region each electron beam splits into two beams, which propagate at different angles and have different chiralities:
\begin{equation}
\label{chi_12}
\chi^{(1)}_{1,2}=\binom{\chi_{1,2}}{1}\;, \qquad \chi_{1,2}=\pm \dfrac{k_y+iK_{1,2}}{\sqrt{k_y^2+K_{1,2}^2}}\,,
\end{equation}
where $K_1$ is $x$-component of the wavevector of the upper mode with positive chirality $\chi_1$ and $K_2$ corresponds to the lower mode with negative chirality $\chi_2$. It is essential that $K_1<K_2$. The incident and refracted beams as well as their spin polarizations are illustrated in Fig.~\ref{birefringence}. The transmitted beam amplitudes are $A_+$ and $B_+$ for the incident beam with positive chirality, and $A_-$ and $B_-$ for negative chirality. The $x$ and $y$ components of the spin polarization in the barrier are estimated as
\begin{equation*}
\begin{split}
S_x\propto (|A_+|^2\!+\!|A_-|^2)\mathrm{Re}\chi_1\!-\!(|B_+|^2\!+\!|B_-|^2)\mathrm{Re}\chi_2\!+ \dots ,\\
S_y\propto (|A_+|^2\!+\!|A_-|^2)\mathrm{Im}\chi_1\!-\!(|B_+|^2\!+\!|B_-|^2)\mathrm{Im}\chi_2\!+ \dots\,,
\end{split}
\end{equation*}
where the dots denote spatially dependent terms originating from the interference.

The total spin polarization is determined by the sum over all incident angles. To estimate this sum, consider the dependence of the spin components on $k_y$. It is clear that $|A_{\pm}|^2$ and $|B_{\pm}|^2$ are even functions of $k_y$. Real and imaginary parts of $\chi_{1,2}$ are seen from Eq.~(\ref{chi_12}) to be correspondingly odd and even functions of $k_y$. Therefore $S_x$ is an even function of $k_y$ and $S_y$ is an odd function, as it is illustrated in Fig.~\ref{birefringence}. Thus the total $S_x$ component vanishes while $S_y$ is nonzero.

It is easy to find the direction of the total spin for the energy close to the barrier height. If $k_y=0$, the amplitudes $|A_+|$ and $|B_-|$ are equal, $|A_-|=|B_+|=0$ and $\chi_1=\chi_2=i$. Hence, the total spin density is zero because the spins of opposed chiralities cancel each other. At small $k_y$, the spin polarization appears. Since the amplitudes are functions of $k_y^2$, they remain unchanged in the first approximation, so that the resulting spin is determined by the difference $(\chi_1-\chi_2)$ and 
\begin{equation*}
S_y \propto \dfrac{K_1}{\sqrt{k_y^2+K_1^2}}-\dfrac{K_2}{\sqrt{k_y^2+K_2^2}}\,. 
\end{equation*}
Since $K_1<K_2$, the $y$-component of the spin density is negative, i.e. the sign of the spin polarization is determined by the lower-energy branch of propagating mode.

If the barrier width is finite, there are four modes in the barrier, but the above property remains unchanged. It is also kept for other branches of the electron spectrum.

\section{The case of Dresselhaus SOI}
\label{Dresselhaus}

All the above results are generalized to the case of Dresselhaus SOI. For a 2D system oriented along [001] crystallographic direction, the SOI Hamiltonian is~\cite{Winkler}
\begin{equation*}
H_D=\dfrac{\hbar^{2}}{2m}(p_{x}^{2}+p_{y}^{2})+
\frac{\alpha}{\hbar}(p_{x}\sigma_{x}-p_{y}\sigma_{y})\,.
\label{HamiltD} 
\end{equation*}
It is well known that Dresselhaus and Rashba Hamiltonians are unitary equivalent.~\cite{RashbaSheka} An unitary matrix
\begin{equation}
\label{R-D_matrix}
U= \left( \begin{array}{cc}
           0 & i\\ 1 & 0
          \end{array}
\right)
\end{equation}
transforms the Rashba Hamiltonian~(\ref{HamiltR}) to the Dresselhaus one: $\tilde{H}_R=U^+H_RU=H_D$. Therefore, the Dresselhaus SOI case does not require separate calculations. It is enough to carry out this transformation. Taking into account that matrix (\ref{R-D_matrix}) transforms the spin matrices as follows: $\tilde {\sigma}_x=-\sigma_y$, $\tilde {\sigma}_y=-\sigma_x$, and $\tilde {\sigma}_z=-\sigma_z$, we arrive at the conclusions:
(i) the dispersion equation is the same as in the Rashba case~(\ref{zeta});

(ii) the spin functions differ from those defined by Eq.~(\ref{chi}) by a simple substitution $k_x\leftrightarrows k_y$;

(iii) the spin components of transmitted electrons in the incident states $|k_x,k_y,\uparrow\rangle$ and $|k_x,k_y,\downarrow\rangle$ differ from those of Sec.\ref{Tunneling} by replacements: $\mu_x\to-\mu_y$,  $\mu_y\to-\mu_x$ and $\mu_z\to-\mu_z$;

(iv) the spin polarization of the current transmitted through a barrier with SOI has only $x$ component, if the distribution function is even with respect to the transverse momentum. In particular, Figs~\ref{polar-d_3}, \ref{polar-E} are valid for the spin polarization normal to the barrier, $\mathcal{P}_x$.

Of course, if the Rashba and Dresselhaus mechanisms act simultaneously the results change qualitatively.

\section{Conclusion}
\label{Conclude}
We have found the total spectrum of electron states in a bounded 2D electron gas with SOI. It addition to well known propagating states it contains two branches of evanescent states. Their wavefunctions decay with the distance in the direction $x$ perpendicular to the boundary. One branch (``purely decaying'' evanescent mode) is described by an imaginary wavevector. The energy of this state is splitted by spin, so that there are two subbranches. They fill in the gap, which opens in the propagating state spectrum at $k_y\ne0$. Other branch (``oscillating'' evanescent mode) is characterized by a complex wavevector $K_x$. These states lie in the forbidden gap.

We have studied the electron transmission through a lateral potential barrier with the SOI. In the energy range, where electrons tunnel via the oscillating evanescent states, the tunneling reveals unusual features, such as an oscillatory dependence of the transmission coefficients on the barrier width and the energy. But of most importance is the spin polarization of the electron current. The value and direction of the polarization depend on the angle of incidence and the energy of incident electrons. The polarization appears even if the distribution function of incident electrons is symmetric with respect to the transverse momentum. In this case the polarization is directed parallel to the barrier (in the Rashba SOI case) or perpendicular to it (for Dresselhaus SOI). The highest efficiency of the spin polarization is attained when the Fermi energy is close to the barrier height. In this case, electrons pass through the barrier partially via the propagating states and partially via the purely decaying evanescent states. Under this condition the most effective spin filtering occurs. The maximal polarization efficiency depends on the barrier height and can exceed 0.5 if the barrier width is on the order of $\pi/a$.

\acknowledgments
This work was supported by Russian Foundation for Basic Research (project No. 05-02-16854), Russian Academy of Sciences (programs ``Quantum Nanostructures''and ``Strongly Correlated Electrons in Semiconductors, Metals, Superconductors, and Magnetic Materials''), RF Ministry of Education and Science.

\end{document}